\documentclass[preprint,showpacs,amsmath,amssymb,prl]{revtex4}
\usepackage{dcolumn}% Align table columns on decimal point
\usepackage{bm}% bold math   
\usepackage{psfig}% bold math
\begin{document}

\preprint{Draft-PRL}

\title{\large\bf Study of the $J/\psi$ decays to 
$\Lambda\overline{\Lambda}$ and
$\Sigma^0\overline{\Sigma}^0$} 
\author{
M.~Ablikim$^{1}$,              J.~Z.~Bai$^{1}$,               Y.~Ban$^{11}$,
J.~G.~Bian$^{1}$,              X.~Cai$^{1}$,                  H.~F.~Chen$^{16}$,
H.~S.~Chen$^{1}$,              H.~X.~Chen$^{1}$,              J.~C.~Chen$^{1}$,
Jin~Chen$^{1}$
% ...
\\
\vspace{0.2cm}
(BES Collaboration)\\
\vspace{0.2cm}
{\it
$^{1}$ Institute of High Energy Physics, Beijing 100049, People's Republic of China\\
% ...
$^{11}$ Peking University, Beijing 100871, People's Republic of China\\
% ...
$^{16}$ University of Science and Technology of China, Hefei 230026, People's Republic of China\\
% ...
}
}

\date{\today}

\begin{abstract}
 With 58 million produced $J/\psi$ events collected by the BES-II
 detector at the BEPC,  the decays
$J/\psi\rightarrow\Lambda\overline{\Lambda}$ and 
$\Sigma^0\overline{\Sigma}^0$
 are analysed. The branching ratios are measured to be
 $Br(J/\psi\rightarrow\Lambda\overline{\Lambda})=(2.05\pm0.03\pm 0.11
 )\times10^{-3}$ and 
$Br(J/\psi\rightarrow\Sigma^0\overline{\Sigma}^0)=(1.40\pm0.03\pm0.07
)\times10^{-3}$. The angular distribution
is of the form $\frac{\displaystyle dN}{\displaystyle dcos\theta}
=N_0(1+\alpha cos^2\theta)$, with $\alpha=0.65\pm0.12\pm 0.08$
 for $J/\psi\rightarrow\Lambda\overline{\Lambda}$ and 
$\alpha=-0.22\pm0.17\pm 0.09$
for $J/\psi\rightarrow\Sigma^0\overline{\Sigma}^0,$ respectively.
\end{abstract}

\pacs{13.20.Gv, 14.20.Jn, 23.20.En} 

\maketitle

As is well-known from the helicity formalism,
the angular distribution of $B$ in the decay of a neutral
vector resonance $V$ into a baryon-antibaryon pair $B\overline{B}$
is given by~\cite{1} 
$$
\frac{\displaystyle dN}{\displaystyle dcos\theta}\sim
 1+\alpha cos^2\theta,
$$
where $\theta$ is the emission polar angle of $B$ in the $V$ rest 
frame.
The first order calculations~\cite{2}~\cite{3} of the perturbative QCD 
predict the
theoretical value of $\alpha$ at $J/\psi$ energy. 
Table 1 summarizes the theoretical
predictions of $\alpha$
for the decays $J/\psi\rightarrow \Lambda\overline{\Lambda},
\Sigma^0\overline{\Sigma}^0$.
Several  experiments
have measured $\alpha$  for $J/\psi\rightarrow
p\overline{p}$~\cite{4}~\cite{5}~\cite{6}~\cite{7},
$\Lambda\overline{\Lambda},
\Sigma^0\overline{\Sigma}^0$~\cite{6}~\cite{7}
and $\Xi^-\overline{\Xi}^+$~\cite{6}.

     In previous papers~\cite{8}~\cite{9}, the analyses of 
$J/\psi\rightarrow 
\Lambda\overline{\Lambda}$ and  $\Sigma^0\overline{\Sigma}^0$ 
using $7.8\times 10^6 J/\psi$ events
collected with the BES-I detector have been reported. The BES-I
 value of
$\Lambda\overline{\Lambda}$ angular distribution coefficient
$\alpha$  
is in good agreement with the DM2 value~\cite{7} as well as the 
theoretical predictions in Ref.[3]. However the BES-I value of
$\Sigma^0\overline{\Sigma}^0$
angular distribution coefficient $\alpha$ with minus sign
 is obviously deviated from 
those of DM2 , MARKII as well as the theoretical prediction.

 This letter presents the analyses of the decays 
$J/\psi\rightarrow\Lambda\overline{\Lambda}$
and $\Sigma^0\overline{\Sigma}^0$ based on  BES-II $58\times 10^6$ 
$J/\psi$ events,
with the purpose of improving the accuracy of branching ratio and
$\alpha$ value measurements and
clarifying  the sign of the $\alpha$ for 
$J/\psi\rightarrow\Sigma^0\overline{\Sigma}^0$.

\begin{center}
Table 1{\hskip 0.4cm} theoretical expectations for $\alpha$
in $J/\psi\rightarrow B\overline{B}.$
{\vskip 0.2cm}
\begin{tabular}{|c|c|c|}
\hline
Channel& Ref.[2]&Ref.[3]\\
\hline
$\Lambda\overline{\Lambda}$ & 0.32&0.51\\
\hline
$\Sigma^0\overline{\Sigma}^0$ &0.31 &0.43\\
\hline
\end{tabular}
\end{center}

{\hskip 0.6cm}The BES-II detector  has been  described in detail 
 elsewhere~\cite{10}.

 {\hskip 0.6cm}Since the decays studied 
 include $\Lambda
\overline{\Lambda}$ pair in the final state, selection criteria are
used to select $J/\psi\rightarrow\Lambda
\overline{\Lambda}+X$ events, where X is a system of neutral
particle(s) and/or undetected charged particle(s).
The $\Lambda$ is identified by its $p\pi^-$ decay mode.
Candidates for 
 $J/\psi\rightarrow\Lambda\overline{\Lambda}+X$ events are
selected by requiring exactly four reconstructed 
charged tracks in the drift
chamber with  zero net charge. 
Tracks with
$\mid cos\theta_{ch} \mid<0.8$ and  transverse
momemtum $p_{xy}>0.07$ GeV 
 are accepted,
where $\theta_{ch}$ is the polar angle
with respect to the beam direction. 
Because the $\Lambda$ is produced at the second vertex,
no limits are required for the primary vertex position.
The particles are
identified by  requiring that their combination weights of
the time-of-flight (TOF) and the ionization energy loss (dE/d{\it x})
in the drift chamber
be consistent with the corresponding particle  hypothesis. 

    The events with four charged particles satisfying
hypothesis of a pair of $p\overline{p}$ and a pair of 
$\pi^+\pi^-$ are selected.    
Fig. 1 shows $p\pi^-$
invariant mass distribution  for  remaining
$J/\psi\rightarrow p\pi^-\overline{p}\pi^+ +X$
candidates (solid line).
By fitting the $M_{p \pi^-}$ distribution  
to a gaussian distribution plus a quadratic polynomial, a 2.6 MeV  
mass resolution is
estimated for the peak.  Fig. 1 also shows 
$p\pi^-$  invariant mass for Monte Carlo simulation (dashed line)
with the mass resolution of 2.6 MeV. 
 To select $\Lambda$ and  $\overline{\Lambda}$,
the requirements of  $\mid M_{p \pi^-}-1.1156 \mid <0.008$ GeV and 
$\mid M_{\overline{p} \pi^+}-1.1156\mid<0.008$ GeV are imposed.

{\hskip 0.6cm}The energy distribution of 
the $\Lambda\overline{\Lambda}$  pair,
$E_{\Lambda\overline{\Lambda}}$ ,for 
 $J/\psi\rightarrow\Lambda\overline{\Lambda}+X$
candidates  is shown in Fig. 2, where 
$E_{\Lambda\overline{\Lambda}}$ is defined as the energy sum of
$\Lambda$ pair,
data is represented by solid line and Monte Carlo 
simulation by dashed line. 

The contamination from non-$\Lambda\overline{\Lambda}+X$
events in Fig. 2 can be estimated by the contribution
from the sidebands, namely the off-$\Lambda$ resonance background 
shown in Fig. 1.
The events in the sidebands are defined as
those in the  squre area of $\mid 
M_{p\pi^-}-1.1156\mid<\sqrt{2}\times3\sigma$
and $\mid M_{\overline{p}\pi^+}-1.1156\mid<\sqrt{2}\times 3\sigma$
minus the  squre area of $\mid 
M_{p\pi^-}-1.1156\mid<3\sigma$ and
$\mid M_{\overline{p}\pi^+-}-1.1156\mid<3\sigma$
in two dimentional $M_{p\pi^-}$ vs. $M_{\overline{p}\pi^+}$ plot, where
$\sigma=0.0026$ GeV is the mass resolution of $\Lambda$ 
$(\overline{\Lambda})$. In Fig. 2 the sideband contribution
is indicated by shaded area.

\begin{center}
\centerline{
\psfig{file=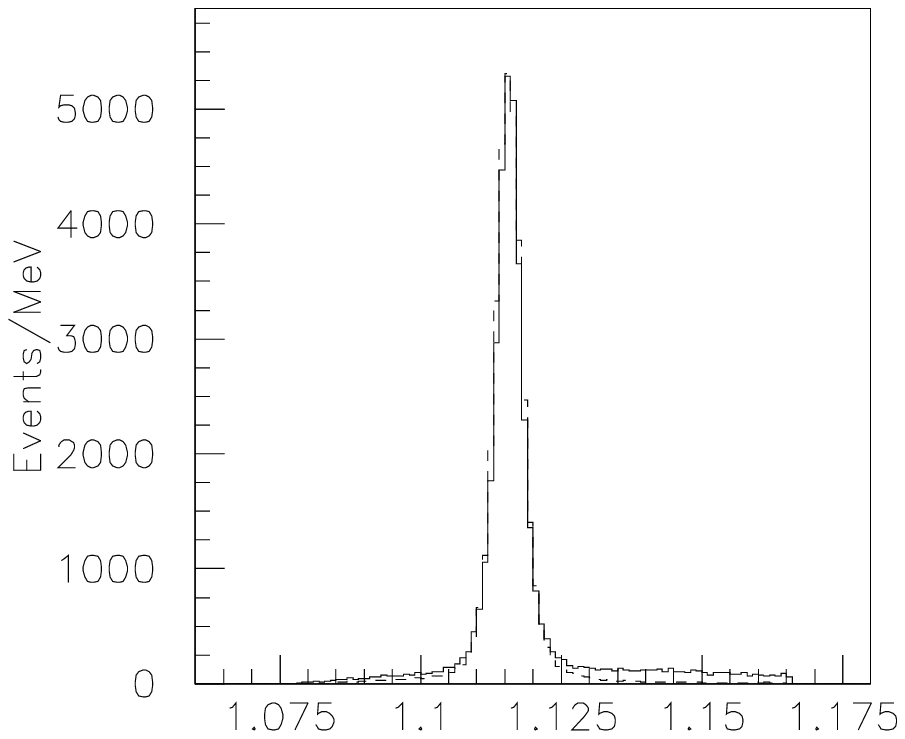 ,bbllx=25pt,bblly=418pt,%
         bburx=272pt,bbury=646pt,width=8cm,height=6cm,clip=}}
\end{center}  
\begin{center}
\vspace{-1.6cm} 
\mbox{}{\hskip 1.2cm} $M_{p\pi^-}$ (GeV)
\end{center}
\vspace{-0.2 cm}
 {\hskip 0.6cm}Fig. 1. $p\pi^-$ invariant mass distribution 
 in  $J/\psi\rightarrow
 p\pi^-\overline{p}\pi^++X$ candidates for data (solid line). 
 The dashed line is Monte Carlo simulation for
 $J/\psi\rightarrow \Lambda\overline{\Lambda}.$
\\
\\
\begin{center}
\centerline{
\psfig{file=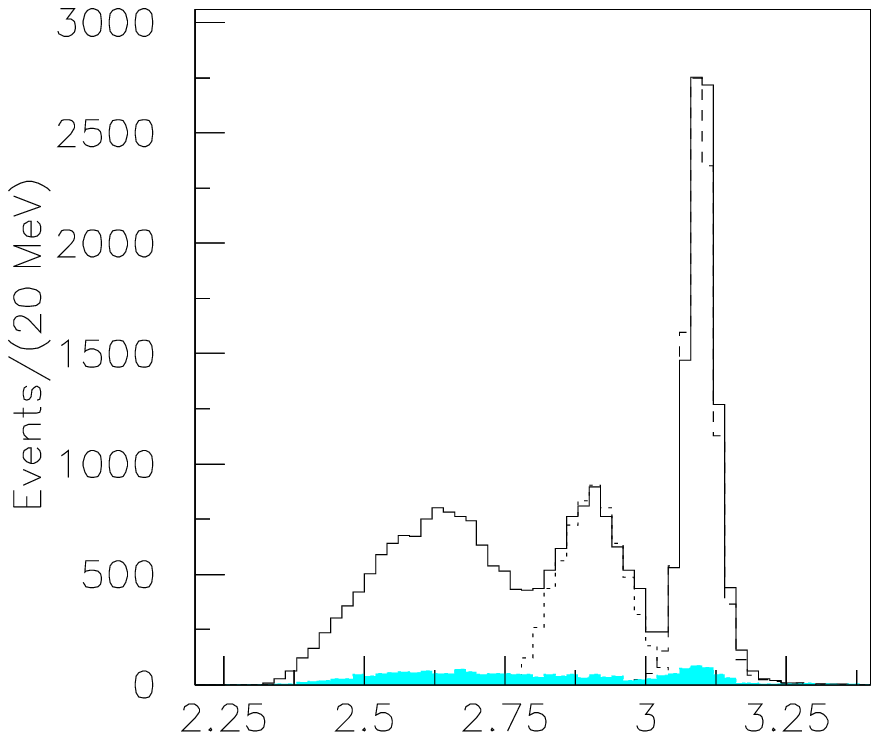 ,bbllx=25pt,bblly=418pt,%
         bburx=272pt,bbury=646pt,width=8cm,height=6cm,clip=}}
\end{center}
\begin{center}
\vspace{-1.6cm}
\mbox{}{\hskip 1.2cm}$E_{\Lambda\overline{\Lambda}}$ (GeV)
\end{center}
\vspace{-0.2 cm}
{\hskip 0.6cm} Fig. 2. Energy distribution of 
$\Lambda\overline{\Lambda}$ for $J/\psi\rightarrow
\Lambda\overline{\Lambda}+X$ candidates. The shaded area is  the
sideband contamination described in the text.
The dashed lines are from Monte Carlo simulation of
$J/\psi\rightarrow \Sigma^0\overline{\Sigma}^0$ and
$J/\psi\rightarrow \Lambda\overline{\Lambda}$.
\\

 A clear  peak centered at
 the $J/\psi $ mass in Fig. 2 from the decay
 $J/\psi\rightarrow\Lambda\overline{\Lambda}$ is observed.
 The enhancement centered at 2.9 GeV  is due to the decay
 $J/\psi\rightarrow\Sigma^0\overline{\Sigma}^0$,where $\Sigma^0
 \rightarrow\Lambda\gamma$. 
  To obtain $J/\psi\rightarrow \Lambda\overline{\Lambda}$
  from the selected $J/\psi\rightarrow\Lambda\overline{\Lambda}
  +X$ candidates,
 contamination from $J/\psi\rightarrow\Sigma^0\overline{\Sigma}^0$
 decay is suppressed by imposing 
 the total missing momentum cut $p_{miss}<0.15$ GeV
 and the $\Lambda$ $(\overline{\Lambda})$
 momentum requirement $0.98<p_{\Lambda}$ 
$(p_{\overline{\Lambda}})<1.2$ 
 GeV.
 Finally by requiring  $3.02\leq 
 E_{\Lambda\overline{\Lambda}}\leq3.2$ GeV a
  $\Lambda\overline{\Lambda}$ sample of
 9462-503=8959 events is obtained,
 here 503 events are from the sidebands,
 which are obtained  
using the same criteria for $\Lambda\overline{\Lambda}$ selection
to the aforementioned non-$\Lambda\overline{\Lambda}+X$ sideband 
events.

   The detection efficiency for the events
 $J/\psi\rightarrow\Lambda\overline{\Lambda}\rightarrow
 p\pi^-\overline{p}\pi^+$ depends on the $\Lambda$
 direction. A Monte Carlo simulation
 gives the detection efficiency $\epsilon(\theta_i)$ 
 in different $\Lambda$ polar angle $\theta_i$ with the
 bin size $\Delta cos\theta_i=0.1 $.
The efficiency corrected angular distribution of
$\Lambda$ for  the decay
$J/\psi\rightarrow\Lambda\overline{\Lambda}$  
 is shown as histogram line in Fig. 3.
   Fitting this angular distribution to the
 theoretical form
  $$\frac{\displaystyle dN}{\displaystyle dcos\theta}=
   N_0(1+\alpha cos^2\theta),$$
yields
$$N_0=(1990.9\pm40.1)/0.1,$$
$$\alpha=0.65\pm0.12,$$
where the  error is statistical. 

Using the $N_0$ and $\alpha$ values, the number of
$J/\psi\rightarrow\Lambda\overline{\Lambda}$ events
corrected for the efficiency $\epsilon_i$ is deduced from
the equation $N_{\Lambda\overline{\Lambda}}=\int^1_{-1}N_0(1+\alpha 
cos^2\theta)dcos\theta,$
and the branching ratio is obtained by the equation
$$Br(J/\psi\rightarrow\Lambda\overline{\Lambda})=
\frac{\displaystyle N_{\Lambda\overline{\Lambda}}}{\displaystyle
Br^2(\Lambda\rightarrow p\pi^-)
 N_{J/\psi}},$$
where 
$N_{J/\psi}=58\times{10^6}$ is the number of $J/\psi$ events
with the error of $4.72\%$~\cite{11}.

The remained contamination from 
$J/\psi\rightarrow\Sigma^0\overline{\Sigma}^0$ is $0.8\%$ 
using the branching ratio determined by this work. 
 Subtracting this contamination 
the number of $J/\psi\rightarrow
\Lambda\overline{\Lambda}$ signal events is 8887.
The branching ratio and angular distribution coefficient are obtained 
to be
$$Br(J/\psi\rightarrow\Lambda\overline{\Lambda})
=(2.05\pm 0.03\pm 0.11)\times10^{-3},$$
$$\alpha=0.65\pm0.12\pm0.08,$$
where the first error is statistical and the second error is 
systematic. The systematic error includes the uncertainty of the 
detection efficiency due to imperfection of Monte Carlo simulation,
the contamination from the indefinite branching ratio of background channel 
$J/\psi\rightarrow \Sigma^0\overline{\Lambda}+c.c.$~\cite{12} and 
the uncertainty of the number of $J/\psi$ events.

\begin{center}
\centerline{
\psfig{file=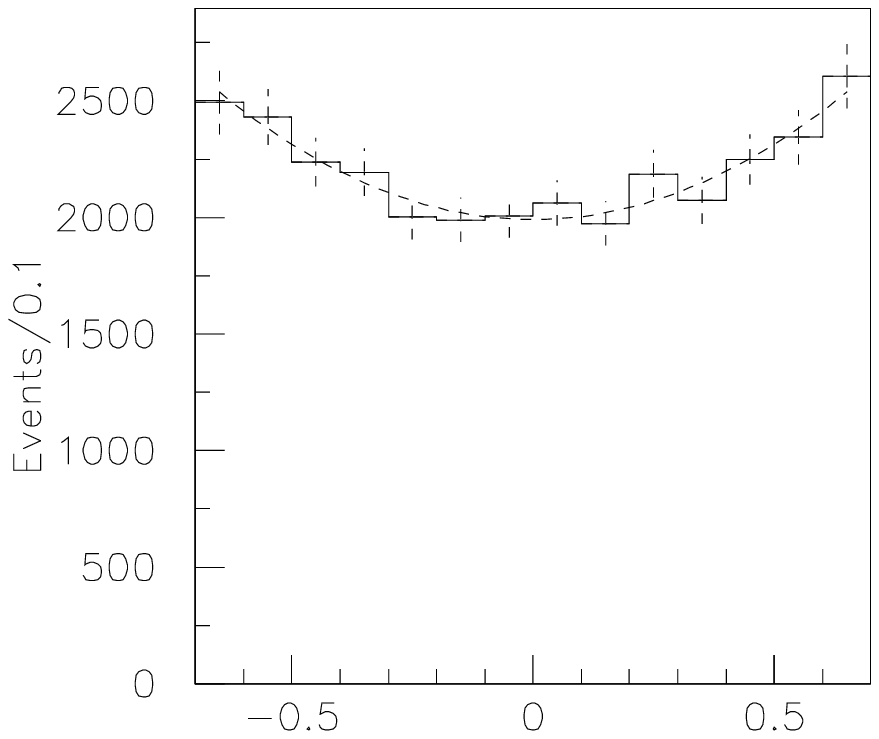 ,bbllx=25pt,bblly=418pt,%
         bburx=272pt,bbury=646pt,width=8cm,height=6cm,clip=}}
\end{center}
\begin{center}
\vspace{-1.6cm}
\mbox{}{\hskip 1.2cm} $cos \theta$
\end{center}
\vspace{-0.2 cm}
{\hskip 0.6cm} Fig. 3. The $\Lambda$ angular distribution
 for  $J/\psi\rightarrow\Lambda\overline{\Lambda}$ events.
 The histogram is efficiency corrected data. 
 The smooth dashed line is the fitting result.
  \\

The enhancement centered at 2.9 GeV in Fig. 2 
is dominantly due to the decay
 $J/\psi\rightarrow\Sigma^0\overline{\Sigma}^0$,where $\Sigma^0
 \rightarrow\Lambda\gamma$. 
$2.75<E_{\Lambda\overline{\Lambda}}<3.02$ is required 
to select $\Sigma^0\overline{\Sigma}^0$ events.
To remove the background from 
$J/\psi\rightarrow\Lambda\overline{\Lambda}$, 
$\Lambda\overline{\Lambda}\pi^0$,
$\Xi^0\overline{\Xi}^0$,
$\Lambda\overline{\Lambda}\gamma$ and
$\Sigma^0\overline{\Lambda}+c.c.$,
 the events are kinematically fitted
to the $J/\psi\rightarrow\Lambda\overline{\Lambda}\gamma\gamma$
topology by imposing
energy and momentum constraints (4C).
The combination with the smallest $\chi^2$ in the 4C fit is chosen to
identify the radiative photons if there are more than two
photon candidates in an event, here a photon is defined as
a cluster with deposite energy larger than 30 MeV in the barrel 
shower counter,
outside a $25^\circ$ cone around $\overline{p}$ and outside a 
$12^\circ$ cone around the other charged particles.  
Then a 6c fit ( two additional constraints $M_{\Lambda\gamma}
=M_{\Sigma^0}$, 
$M_{\overline{\Lambda}\gamma}=M_{\overline{\Sigma}^0}$)
is performed. It is required that
$\chi^2_{\Lambda\overline{\Lambda}\gamma\gamma}(4c)<
\chi^2_{\Lambda\overline{\Lambda}\gamma}(4c).$ 
The $M_{\overline{\Lambda}\gamma}$ distribution is shown in Fig. 4,
where the solid line is data and the dashed line is Monte Carlo
simultaion $J/\psi\rightarrow\Sigma^0\overline{\Sigma}^0.$ The
$M_{\Lambda\gamma}$ distribution is similar to Fig. 4.
Finally, after applying
$\chi^2_{\Lambda\overline{\Lambda}\gamma\gamma}(6c)<40$,
2194-82=2112 events are remained, where 82 is from the sidebands
 which are obtained
using the same criteria for $\Sigma^0\overline{\Sigma}^0$ selection
to the aforementioned non-$\Lambda\overline{\Lambda}+X$ sideband
events.

\begin{center}
Table 2 {\hskip 0.4cm} The background for 
$J/\psi\rightarrow\Sigma^0\overline{\Sigma}^0$
{\vskip 0.2cm}
\begin{tabular}{|c|c|}
\hline
Channel& contamination ($\%$)\\
\hline
$\Lambda\overline{\Lambda}$ & 0.1\\
\hline   
$\Xi^0\overline{\Xi}^0$ &0.3\\
\hline
$\Lambda\overline{\Lambda}\pi^0$&0.5\\
\hline
$\Sigma^0\overline{\Lambda}+c.c.$&$<0.08$\\
\hline
$\Lambda\overline{\Lambda}\gamma$&$<0.08$\\
\hline 
\end{tabular} 
\end{center}

 The detection efficiency for the signal events
 depends on the $\Sigma^0$ direction.
 The  angular dependence of $\Sigma^0\overline{\Sigma}^0$ detection 
efficiency
 is obtained 
 by selecting Monte Carlo events
$J/\psi\rightarrow\Sigma^0\overline{\Sigma}^0
\rightarrow\Lambda\gamma\overline{\Lambda}\gamma
\rightarrow p\pi^-\gamma\overline{p}\pi^+\gamma$
 through the same selection criteria. 
 The efficiency corrected angular distribution
is  shown in Fig. 5 as histogram.
   Fitting this angular distribution to the
 theoretical form
  $$\frac{\displaystyle dN}{\displaystyle dcos\theta}=
   N_0(1+\alpha cos^2\theta),$$
yields
$$N_0=(1786.9\pm57.4)/0.1,$$ 
$$\alpha=-0.22\pm0.17,$$
where $\theta$ is the polar angle between the $\Sigma^0$ and $e^+$
beam and the error is statistical. The number of 
$J/\psi\rightarrow\Sigma^0\overline{\Sigma}^0$
events corrected for the detection efficiency is 
$N_{\Sigma^0\overline{\Sigma}^0}=\int^1_{-1}N_0(1+\alpha 
cos^2\theta)dcos\theta$ and
the branching ratio is calculated with
$$Br(J/\psi\rightarrow\Sigma^0\overline{\Sigma}^0)=
\frac{\displaystyle N_{\Sigma^0\overline{\Sigma}^0}}{\displaystyle 
Br^2(\Lambda\rightarrow p\pi^-)
 N_{J/\psi}}.$$

Table 2 summarizes 
the remaining contamination to $J/\psi\rightarrow 
\Sigma^0\overline{\Sigma}^0$ candidates from  background channels. 
The branching ratio of $J/\psi\rightarrow \Lambda\overline{\Lambda}$
is taken from this work, while the others are from the 
PDG. Subtracting all the contaminations of first three channels, the 
number of $J/\psi\rightarrow \Sigma^0\overline{\Sigma}^0$ 
signal events is 2093. The backgrounds of other two channels are
considered in the systematic error.
The results for the branching ratio of $J/\psi\rightarrow 
\Sigma^0\overline{\Sigma}^0$ and its angular distribution
coefficeincy are
$$Br(J/\psi\rightarrow \Sigma^0\overline{\Sigma}^0)=(1.40\pm0.03\pm 
0.07)\times 10^{-3},$$
$$\alpha=-0.22\pm0.17\pm0.09,$$
respectively, where the first error is statistical and the second 
systematic.  The systematic error includes the uncertainty of 
detection efficiency due to imperfection of Monte Carlo simulation,
the contaminations from the indefinite branching ratios of background 
channels
$J/\psi\rightarrow \Lambda\overline{\Lambda}\gamma$ and
$J/\psi\rightarrow \Sigma^0\overline{\Lambda}+c.c.$ and
 the uncertainty of the number of $J/\psi$ events.

 {\hskip 0.6cm}Table 3 and Table 4 summarize the branching ratios and
$\alpha$ values measured by this work
for the decays $J/\psi\rightarrow\Lambda\overline{\Lambda}$
and $J/\psi\rightarrow\Sigma^0\overline{\Sigma}^0$
together with those previously reported by  Mark II~\cite{6} and 
DM2~\cite{7}.

\begin{center}
\centerline{
\psfig{file=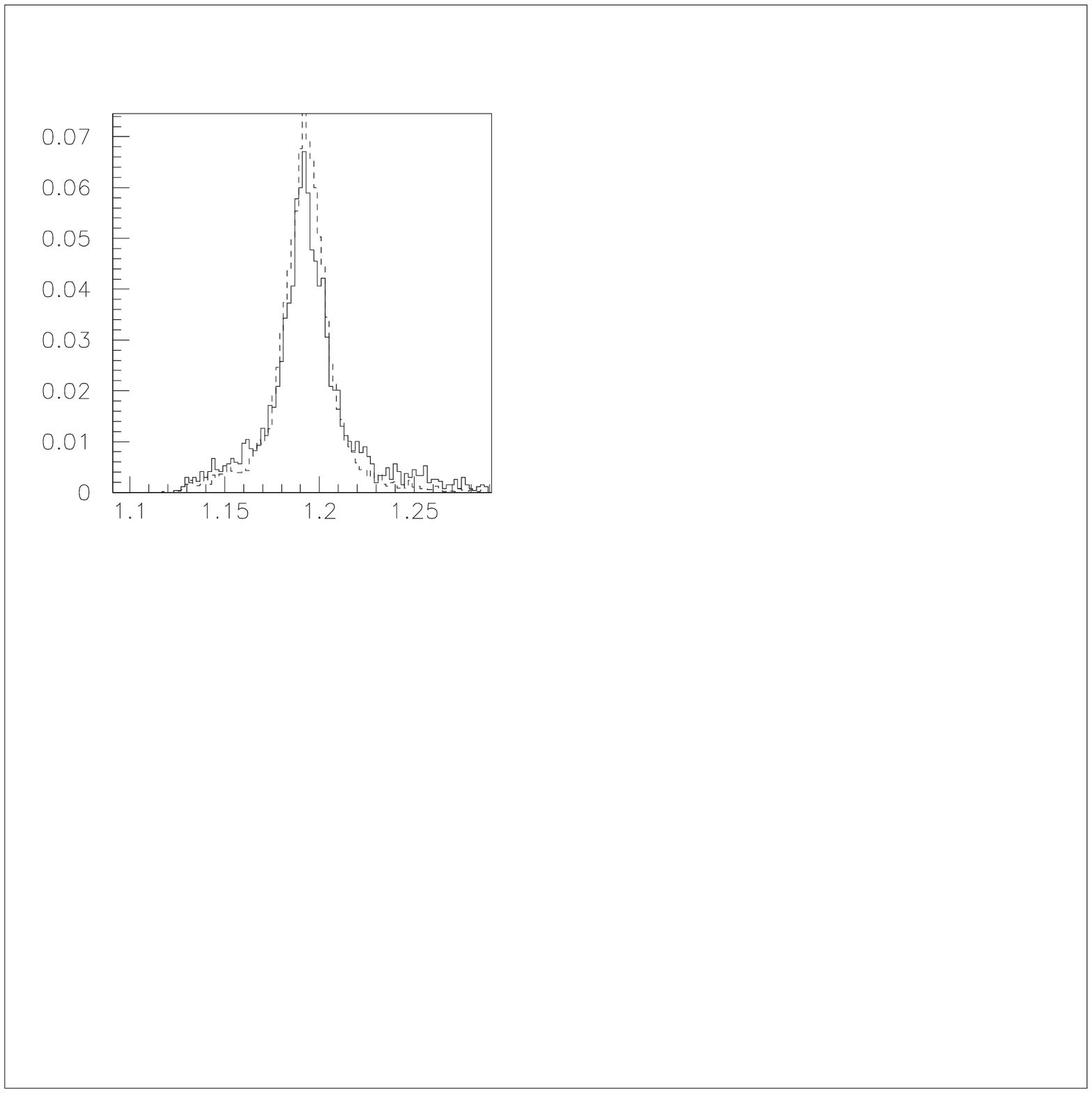 ,bbllx=25pt,bblly=418pt,%
         bburx=272pt,bbury=646pt,width=8cm,height=6cm,clip=}}
\end{center}
\begin{center}
\vspace{-1.6cm}
\mbox{}{\hskip 1.2cm} $M_{\overline{\Lambda}\gamma}$
\end{center}
\vspace{-0.2 cm}
 Fig. 4. The $\overline{\Sigma}^0$ signal after
 4c fit $\chi^2_{\Lambda\overline{\Lambda}}(4c)<40$.
 The solid line is data and the dashed line is Monte Carlo
simultaion $J/\psi\rightarrow\Sigma^0\overline{\Sigma}^0.$

\begin{center}
\centerline{
\psfig{file=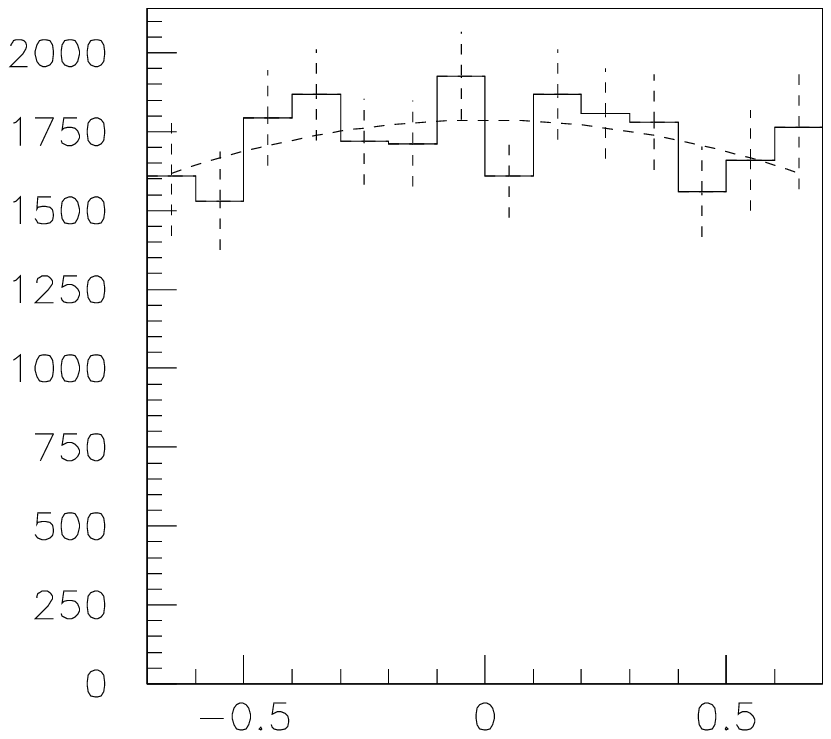 ,bbllx=25pt,bblly=418pt,%
         bburx=272pt,bbury=646pt,width=8cm,height=6cm,clip=}}
\end{center}
\begin{center}
\vspace{-1.6cm}
\mbox{}{\hskip 1.2cm} $cos \theta$
\end{center}
\vspace{-0.2 cm}
 Fig. 5. The $\Sigma^0$ angular distribution
 for $J/\psi\rightarrow\Sigma^0\overline{\Sigma}^0$.
 The histogram is efficiency corrected data. 
The smooth dashed line is the 
fitting result.
\\

 The statistics of the signal events in BES-II
 result is much better than those  of  Mark II, DM2 and BES-I.
 In this experiment the $\alpha$ value for 
$J/\psi\rightarrow\Lambda\overline{\Lambda}$ 
obtained by this experiment is consistent with the previous 
measurements, while the $\alpha$ value is negative for 
$J/\psi\rightarrow\Sigma^0\overline{\Sigma}^0$, which
agrees with that of BES-I~\cite{9} while conflicts
with the values of MARK-II and DM2 and  theoretical 
expectation~\cite{2}~\cite{3}.

 In this analysis the 6c kinematic fit is used to select 
$J/\psi\rightarrow\Sigma^0\overline{\Sigma}^0$ and hence the 
contamination is small; while in DM2~\cite{7} and BES-I 
analyses~\cite{9} the 
$\Sigma^0$ is not reconstructed and
$cos\theta$ of $\Sigma^0$ is replaced with that of $\Lambda$.
This makes $\alpha$ for $J/\psi\rightarrow\Sigma^0\overline{\Sigma}^0$
determined by this work rather believable.

It is worth noting that the central value of $\alpha$ reported by 
DM2~\cite{7} for $J/\psi\rightarrow \Sigma^0\overline{\Sigma}^0$
is positive, however, the angular distribution has a convex shape
as a whole, with the only exception at 
$\cos\theta= -0.65$.   
\\

\begin{center}
Table 3 \\
{\vskip -0.1cm}
Experimental measurements
for decay $J/\psi\rightarrow\Lambda\overline{\Lambda}$\
{\vskip 0.2cm}\
\begin{tabular}{|c|c|c|c|}
\hline
Exp.& $\alpha$ &$Br(\times 10^{-3})$ &Evts\\
\hline
MARKII& $0.72\pm0.36$ & $1.58\pm0.08\pm0.19$ &365\\
\hline
DM2& $0.62\pm0.22$ & $1.38\pm0.05\pm0.20$ &1847\\
\hline
BES-II&$0.65\pm0.14$&$2.05\pm0.03\pm 0.11$&8887\\
\hline
\end{tabular}
\end{center}

\begin{center}
Table 4\\
{\vskip -0.1cm}
 Experimental measurements for 
$J/\psi\rightarrow \Sigma^0\overline{\Sigma}^0$
channel
{\vskip 0.2cm} 
\begin{tabular}{|c|c|c|c|}
\hline
Exp. &$\alpha$& $Br(\times 10^{-3})$& Evts\\
\hline
MARK-II & $0.70\pm1.10$&$1.58\pm0.16\pm0.25$&90\\
\hline
DM2 &$0.22\pm 0.31$&$1.06\pm 0.04\pm0.23$&884\\
\hline
BES-II &$-0.22\pm 0.19$&$1.40\pm 0.03\pm0.07$&2093\\
\hline
\end{tabular}
\end{center}

The BES collaboration thanks the staff of BEPC for their hard
efforts. This work is supported in part by the National Natural
Science Foundation of China under contracts Nos. 10491300,
10225524, 10225525, 10425523, the Chinese Academy of Sciences under
contract No. KJ 95T-03, the 100 Talents Program of CAS under
Contract Nos. U-11, U-24, U-25, and the Knowledge Innovation
Project of CAS under Contract Nos. U-602, U-34 (IHEP), the
National Natural Science Foundation of China under Contract No.
10225522 (Tsinghua University), and the Department of Energy under
Contract No.DE-FG02-04ER41291 (U Hawaii).

\end{document}